# An analysis of the suitability of OpenAlex for bibliometric analyses

Juan Pablo Alperin*, Jason Portenoy**, Kyle Demes**, Vincent Larivière***, and Stefanie Haustein****

*juan@alperin.ca
https://orcid.org/0000-0002-9344-7439
ScholCommLab and School of Publishing, Simon Fraser University, Canada

** jportenoy@ourresearch.org; kyle@ourresearch.org
https://orcid.org/0000-0002-3340-2597; https://orcid.org/0000-0003-2780-0393
OurResearch, USA

***vincent.lariviere@umontreal.ca
https://orcid.org/0000-0002-2733-0689
School of Library and Information Science, University of Montreal, Canada

****stefanie.haustein@uottawa.ca
https://orcid.org/0000-0003-0157-1430
ScholCommLab, School of Information Studies, University of Ottawa, Canada

**Abstract**
Scopus and the Web of Science have been the foundation for research in the science of science even though these traditional databases systematically underrepresent certain disciplines and world regions. In response, new inclusive databases, notably OpenAlex, have emerged. While many studies have begun using OpenAlex as a data source, few critically assess its limitations. This study, conducted in collaboration with the OpenAlex team, addresses this gap by comparing OpenAlex to Scopus across a number of dimensions. The analysis concludes that OpenAlex is a superset of Scopus and can be a reliable alternative for some analyses, particularly at the country level. Despite this, issues of metadata accuracy and completeness show that additional research is needed to fully comprehend and address OpenAlex's limitations. Doing so will be necessary to confidently use OpenAlex across a wider set of analyses, including those that are not at all possible with more constrained databases.

## 1. Introduction

Bibliographic databases are the foundation for basic research in the science of science and inform some of the most important decisions in research management. For many years, proprietary databases such as Scopus and Web of Science have been the only two data sources for these purposes. Decades of analyses have shown that these databases systematically under represent some disciplines and areas of the world, and as a result, present a skewed understanding of the research system (Archambault et al., 2006; Basson et al., 2022; Larivière et al., 2006). However, their continued use over such an extended period has allowed the bibliometrics community an opportunity to understand and document the nature of these databases, their inherent biases, and limitations, further perpetuating their use.

New inclusive databases have emerged in recent years and have made apparent the vast array of scholarship that falls outside the scope of these two canonical databases. Projects like Dimensions and OpenAlex affirm the need to "recalibrate the scope of scholarly publishing" (Khanna et al., 2022). OpenAlex has been seen by many to hold the most promise, not least because of their non-profit nature and their commitment to making all their data and source code openly available (Priem et al., 2022).

A growing number of studies use OpenAlex as a data source (for a list of works that reference OpenAlex, see OpenAlex, n.d.). These include bibliometric (Akbaritabar et al., 2023; Haunschild & Bornmann, 2024) and altmetric studies (Arroyo-Machado & Costas, 2023; Ismail et al., 2023; Mongeon et al., 2023) as well as scoping reviews (Woelfle et al., 2023) and the creation of a knowledge graph for AI innovation (Massri et al., 2023). Many of these studies consider OpenAlex as a convenient new data source; however, they do not critically assess its limitations.

To determine whether OpenAlex can serve as a replacement for the closed databases, its strengths and weaknesses must be carefully documented. Some studies have begun testing OpenAlex's suitability for bibliometric analyses. While these studies find that OpenAlex usually outperforms other databases on coverage (Culbert et al., 2024; Jiao et al., 2023), they also notice issues in the completeness and correctness of bibliographic metadata. For instance, fields such as publication date, volume, issue and page number (Delgado-Quirós & Ortega, 2024), abstracts, author names, references and citations (Culbert et al., 2024; Mongeon et al., 2023), institutions (Akbaritabar et al., 2023), funder information (Schares, 2024) or open access status (Jahn et al., 2023; Simard et al., 2024) have been shown to be incomplete or contain wrong information.

This study, done in collaboration with a team at OpenAlex, can be seen as part of these efforts to assess OpenAlex's suitability for bibliometric analyses. It does so by addressing the following research questions:

RQ1. What is the extent of the coverage found in OpenAlex, in terms of a) document types, b) author affiliation countries, c) number of references, and d) number of citations indexed?

RQ2. Do analyses performed using OpenAlex and Scopus yield comparable results when using their respective metadata?

## 2. Methods

Our analysis used the OpenAlex data snapshot from November 21, 2023. This data is freely available and open for reuse with a CC0 license (OpenAlex, 2024). It was downloaded, processed and analyzed using Python and the Pandas library (Alperin, 2024). We removed works of type "paratext" from our analyses, but all other works are included unless otherwise noted.

Scopus data used was extracted from the scopus.com and scival.com user interfaces. All analyses detailed below were retrieved from exporting the filter counts from Scopus results pages for queries that returned all works for the time windows of interest (all years: "PUBYEAR > 0"; and 2013-2022: "PUBYEAR > 2012 AND PUBYEAR < 2023") on December 1-6, 2023. These filter counts provided all of the Scopus data needed for the study except for citation counts by country and counts for green OA documents that do not belong to other OA categories. To retrieve the green only OA counts in Scopus, we added the following query line to date queries described above: "OA(repository) AND NOT ( OA(publisherfullgold OR publisherhybridgold OR publisherfree2read ) )". Citation counts by country cannot be readily retrieved from Scopus but can be retrieved from SciVal (which uses Scopus data) for the 2013-2022 time period (but not for the full time period, as with other data). We retrieved data on the number of citations in SciVal for each country between 2013-2022 using the Benchmarking module on March 14, 2024.

To compare the overlap between the two databases, we retrieved the full list of sources indexed in Scopus along with identifying metadata (i.e., ISSN) from scopus.com/sources on October 25, 2023. We labelled each work in OpenAlex as "in Scopus" or "not in Scopus" based on whether or not that work had a source that had been matched to a Scopus source using ISSN. In some cases, this led to OpenAlex reporting a higher number of works than Scopus for a given source. The combined "in Scopus" and "not in Scopus" is described as "OpenAlex Total" in the subsequent analysis.

OpenAlex and Scopus classify documents differently (Delgado-Quirós & Ortega, 2024). While both use a similar number of categories (OpenAlex 14 and Scopus 16), there is little overlap between the two systems used. To compare between the two, we re-coded documents into a simplified schema of 9 document types, using the OpenAlex categories as the basis, and considering the distinctions that are typically made in bibliometric analysis. As such, some document types not tracked by Scopus, such as peer reviews and datasets, were grouped into an "other" category and some distinctions made by Scopus, such as conference papers and reviews, were grouped into "article". A mapping of document types is available in the supplementary materials.

To compare the number of works from authors from each country we had to resolve the differences in the way each database classifies territories. The OpenAlex data links works to 247 different countries, whereas Scopus data contains works linked to 239. Of the 248 entities that OpenAlex considers countries, 44 were identified as territories of sovereign nations. This included 15 British territories, 10 French departments and territories, 5 constituent countries of the Kingdom of the Netherlands and 5 US territories. This resulted in a final list of 204 countries (including Antarctica, Western Sahara and Kosovo). Scopus countries were also mapped to this list. The mapping of countries found in the data to the final list of 204 can be found in the supplementary materials.

## 3. Results

*3.1. What is the extent of the coverage found in OpenAlex?*
There are 168.2M works in OpenAlex published between 2000 and 2022, with a peak of 10.2M works published in 2020 (Figure 1). OpenAlex classifies works into 14 types, but the majority (82%) of these works are classified as articles. The proportion classified as articles changes only slightly (to 79%) when considering only the most recent 10 years. To better understand this largest category, we divided it by the source it was published in (i.e., journal, repository, conference, or other). We found that the majority of articles (71%) are found in journals, with only minor fluctuations in the proportions between types in the most recent years. Notably, there is a sizable proportion of articles for which the type of source is not known (lightest blue in Figure 1). The number of articles with an unknown source reduces significantly in 2022, perhaps connected to the closing of Microsoft Academic Graph.

Works can be linked to a country by observing the affiliation of its authors; however, there are a substantial number of works in OpenAlex for which no country can be identified. The USA is the country with the most works, with a growing number of works each year, followed by China, with a fluctuating number of publications each year (Figure 2).

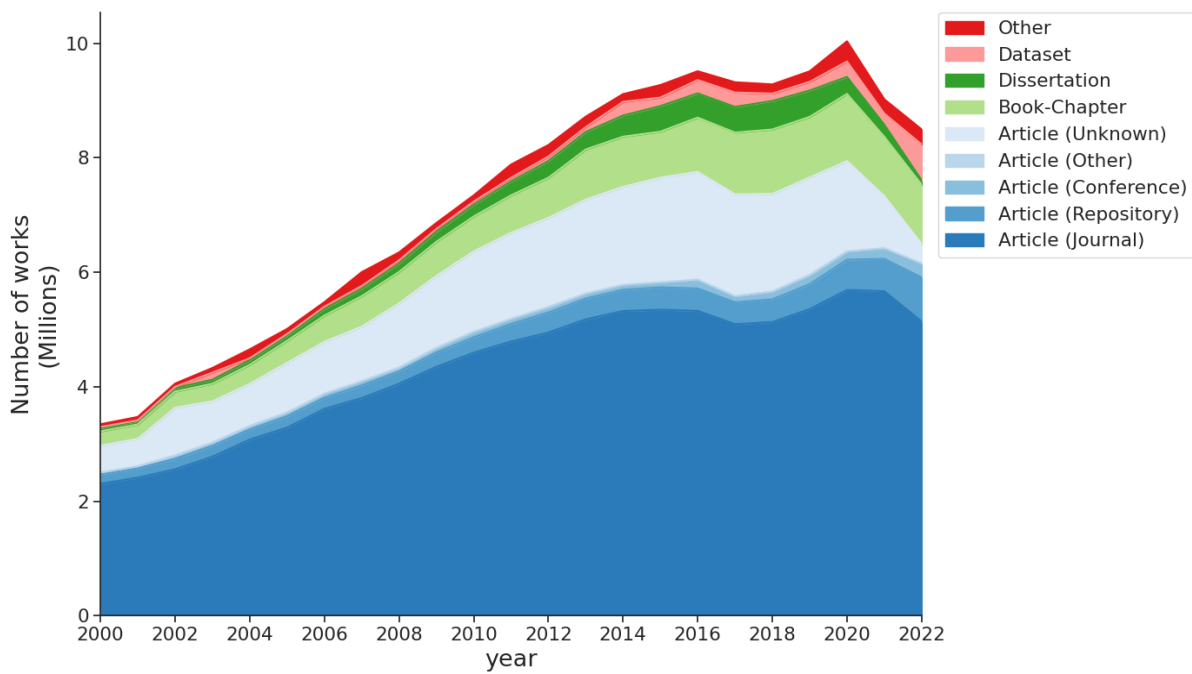

Figure 1: Number of Works in OpenAlex per Year by Document Type.

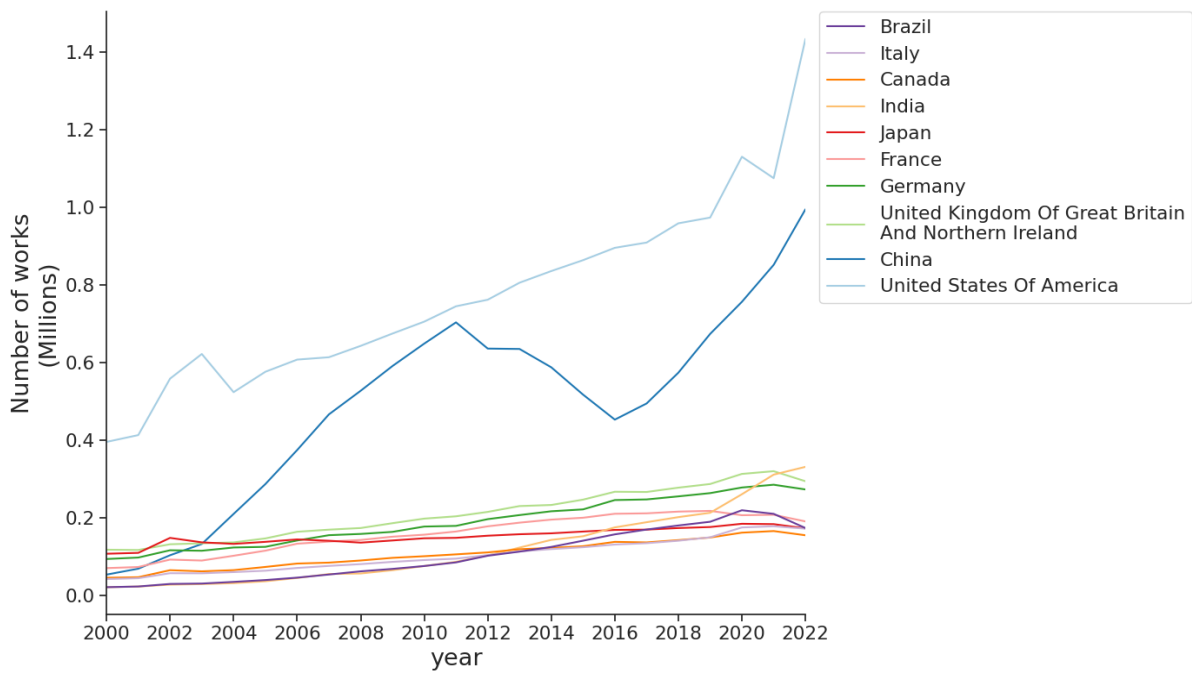

Figure 2: Number of Works in OpenAlex per Year by Country (top 10).

While OpenAlex indexes the metadata of nearly 170M works, it does not contain a complete citation graph for each of these works. We found that two thirds of the works (67%) do not have any references (Figure 3a). While some of those may not actually contain references in their full text, we suspect that for the majority, this is due to the lack of available references for processing. Figure 3a shows the proportion of works with various numbers of references, ranging from 8% of works with 1-5 references to 1% of works containing 36-40 references. Similarly, almost two thirds of works (66%) do not receive any citations (Figure 3b), but this may, again, be because the references have not been indexed and not because the works have not been cited; 16% of works have been cited between 1 and 5 times.

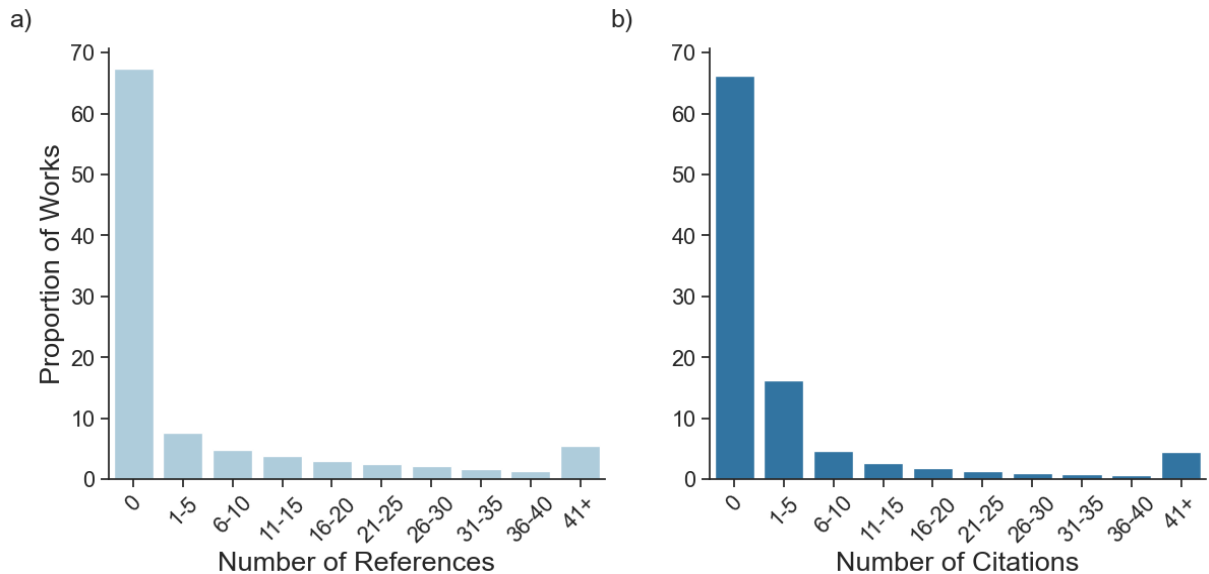

Figure 3: Proportion and Distribution of References and Citations.

*3.2. Matching at the journal level: Does OpenAlex have 100% of Scopus?*

Scopus has 42,948 unique journals with at least one ISSN. Of these, 35,814 (83%) can be matched to an ISSN from a source in OpenAlex. However, Scopus has only 28,085 sources that are considered to be "active," of which 26,513 (94%) can be matched to a source in OpenAlex. While it may seem that OpenAlex is missing Scopus journals, a random spot-check of sources not found in OpenAlex reveals that over half of them are actually present, but missing or erroneous metadata prevented matching them to the source.

As a result of this matching, 88.6M (37%) of the works in OpenAlex works are labelled "in Scopus" and 153.6M (63%) are "not in Scopus." For the more recent years (2013–2022), 29.9M (32%) of the works are "in Scopus" and 63.7M (68%) are "not in Scopus".

*3.3. Do analyses performed using OpenAlex and Scopus yield comparable results?*

We performed analyses to determine whether OpenAlex yields results that are comparable with those obtained using Scopus. Figure 4 shows the Open Access (OA) status in Scopus (orange) compared with the subset of works in OpenAlex whose source was matched to a source in Scopus (dark blue), as well as to the whole of OpenAlex (dark and light blue). We observe that the proportion of works between the various types of OA are largely comparable, with the exception of the Hybrid category, where OpenAlex identifies 81% more works for the "in Scopus" set across all years (58% for the most recent ten years), relative to the number of works in that category in Scopus. For both time periods, Scopus classifies more works as Closed than OpenAlex for the comparable subset. When considering all of OpenAlex, OpenAlex nearly doubles (and in some cases more than triples) the number of works found by Scopus in each category.

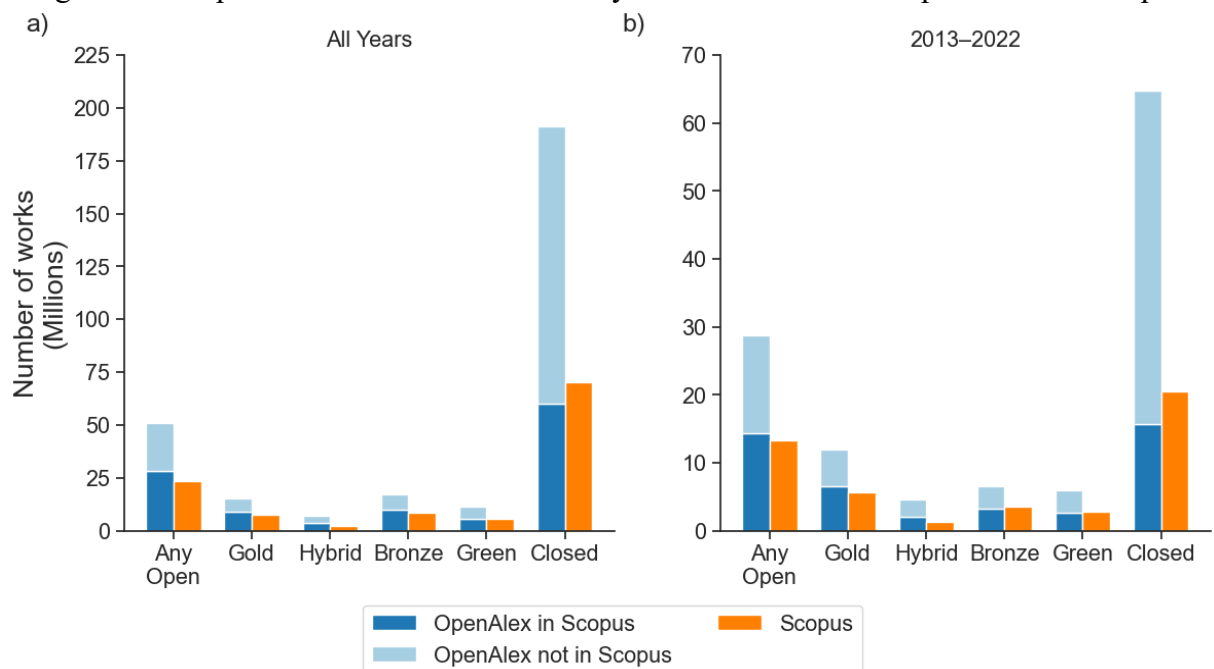

Figure 4: Comparison of Number of Works by OA Status Between OpenAlex and Scopus.

When considering the subset of works in OpenAlex whose source was matched to a source in Scopus, we see that there is little variation in the number of documents classified as articles between both databases. This is true for the most recent ten years, where we see a 4% difference (Table 1) and across all years, where the difference drops to 0% (Supl Table 1). Differences across other document types vary more substantially and seem to reflect the different policies and approaches for indexing.

Table 1: Comparison of Number of Works by Document Type Between OpenAlex and Scopus (2013-2022)

| Document Type | Scopus | OpenAlex | | | |
|---|---|---|---|---|---|
| | | Source In Scopus | Difference | All | Difference |
| article | 30,360,256 | 29,032,344 | -4% | 74,318,093 | 145% |
| book-chapter | 1,788,040 | 712,625 | -60% | 10,001,177 | 459% |
| other | 1,063 | 21,196 | 1894% | 7,564,981 | 711563% |
| book | 198,963 | 31,055 | -84% | 1,342,100 | 575% |
| report | 6 | 4 | -33% | 203,869 | 3397717% |
| editorial | 711,944 | 64,929 | -91% | 101,525 | -86% |
| erratum | 252,202 | 54,691 | -78% | 60,860 | -76% |
| letter | 513,253 | 13,098 | -97% | 17,231 | -97% |
| retracted | 14,788 | 0 | -100% | 0 | -100% |

After mapping countries found in author affiliations in both databases, we observed that works in OpenAlex were linked to two independent countries which do not have any records in Scopus (Kosovo and Bhutan). The inverse (countries in Scopus that were not found in OpenAlex) only occurs for Yugoslavia. After removing these three countries and those works for which no country could be identified, it is clear that there is a high degree of correlation between Scopus and OpenAlex in the number of works per country (Figure 5). Spearman correlations confirm that both the datasets produce similarly ranked lists of countries for the most recent ten years (Table 2) and for all years (Supl. Table 2). In all cases, Spearman $\rho > .95$.

A different representation of the country data, grouped by the World Bank country income groups, shows the differences between the works from OpenAlex whose source was matched to a source in Scopus and those that were not (Figure 6). In addition to revealing the additional works found by OpenAlex (in light blue), Figure 6 shows the large number of works in OpenAlex for which a country was not identified (over 50M works in the most recent 10 years).

Table 2. Spearman Correlation of Countries between OpenAlex and Scopus (2013-2022)

|                    | **OpenAlex in Scopus** | **Scopus** |
|--------------------|------------------------|------------|
| **Scopus**         | 0.989                  | -          |
| **OpenAlex Total** | 0.994                  | 0.983      |

Figure 5: Comparison of Number of Works by Country Between OpenAlex and Scopus

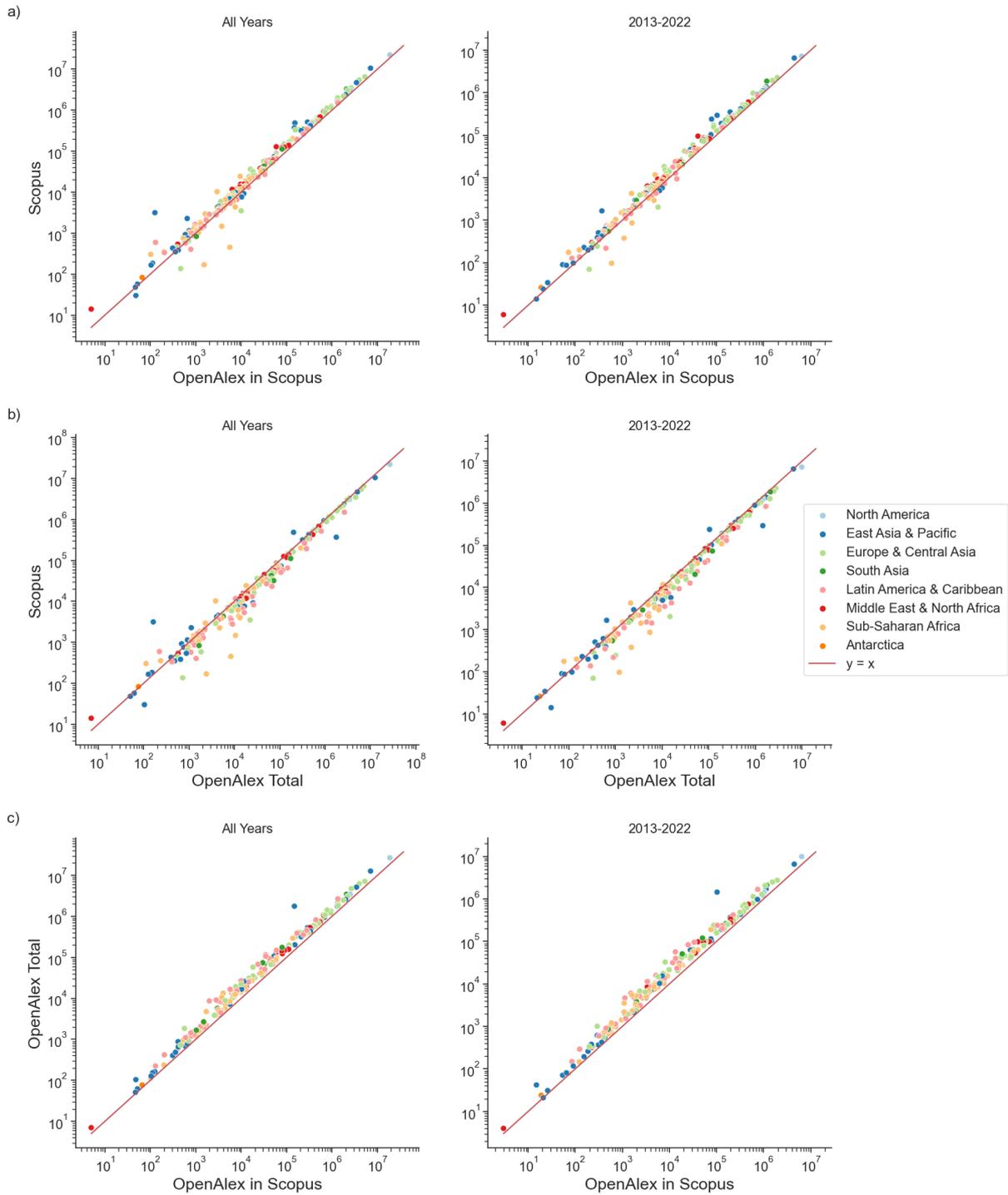

*Figure 6: Comparison of Number of Works by Country Income Group Between OpenAlex and Scopus*

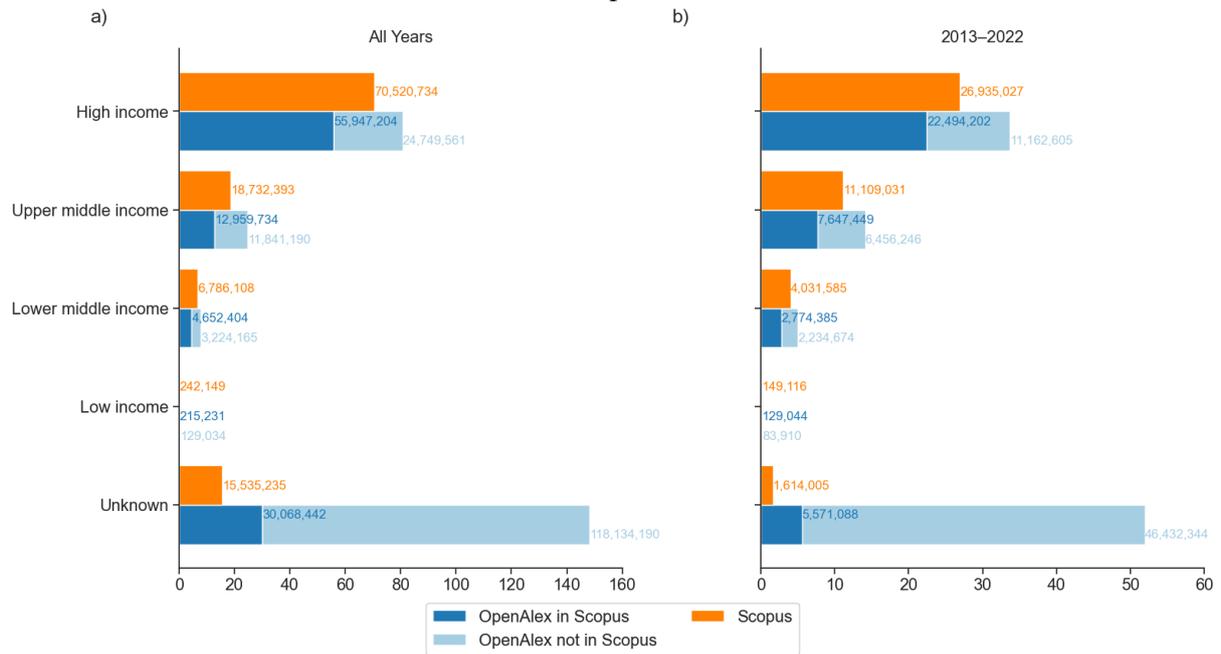

We used the same country mapping to compare the number of citations received by the works in the two databases for the 2013-2022 period. We find that the Spearman correlations between the sets is also very high. All three correlations produce a ρ > .95 (Table 3). When plotting the number of citations by country for the various sets, we observe two related things: first, for almost all countries, Scopus has more citations than OpenAlex, even when considering all the works in OpenAlex that are in and those that are not in Scopus (Figure 7a,b); second, the number of citations per country barely increases when comparing the works in OpenAlex that are in Scopus with all of OpenAlex (Figure 7c).

Table 3. Spearman Correlation of Citations by Country between OpenAlex and Scopus (2013-2022)

|  | **OpenAlex in Scopus** | **Scopus** |
|---|---|---|
| **Scopus** | 0.957 | - |
| **OpenAlex Total** | 1.000 | 0.956 |

Figure 7: Comparison of Citations by Countries and Region.

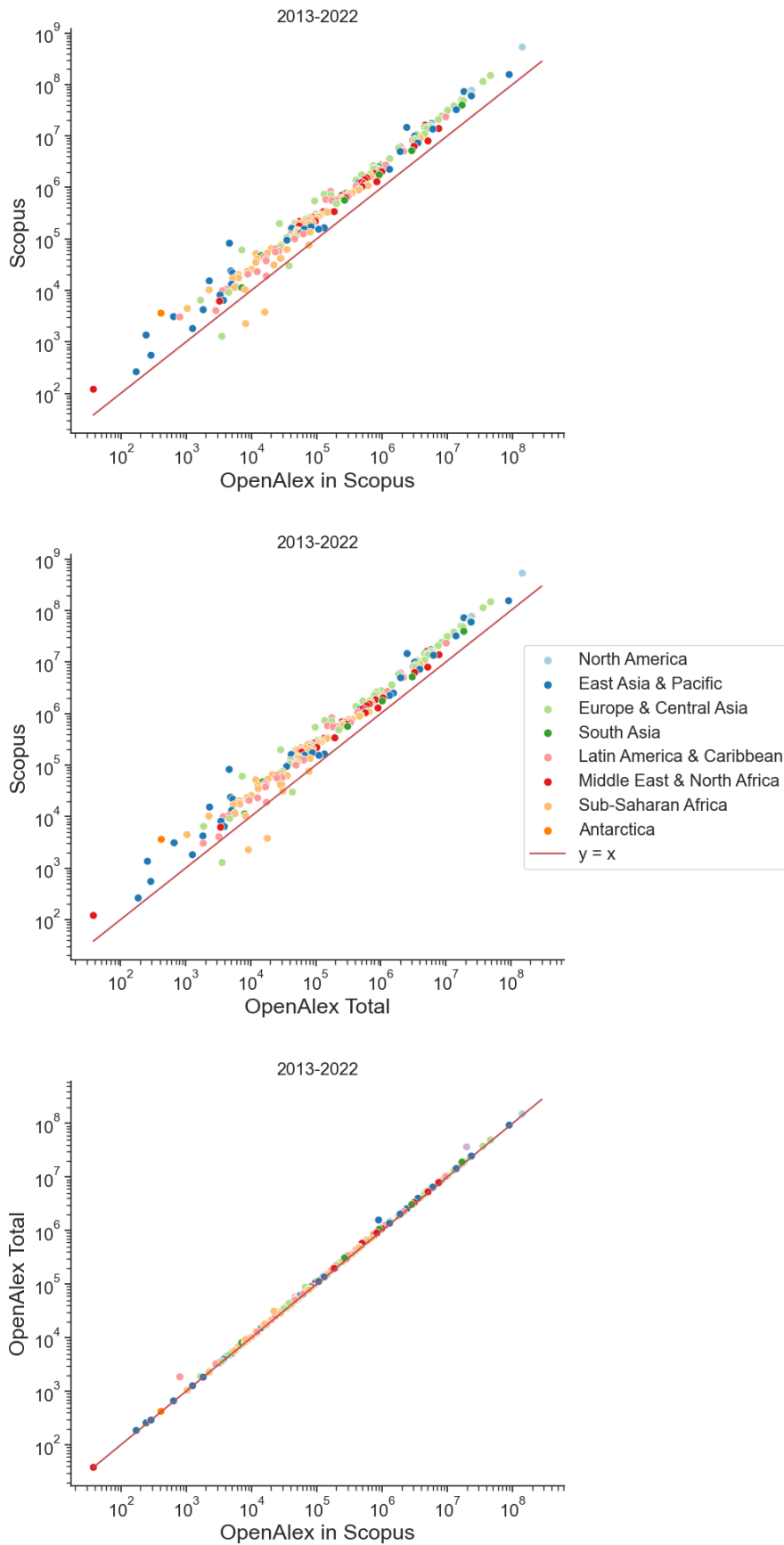

Finally, we compared the number of works by languages following a similar procedure. As with countries, OpenAlex has listed languages not found in Scopus, and vice versa. Languages with more than a thousand works in Scopus, but not found in OpenAlex, are: Bosnian, Icelandic and Serbian; Languages with more than a thousand works in OpenAlex, but not found in Scopus are: Somali, Swahili, and Tamil. Without removing these zeros, the rank correlation between the number of works per language between the three groups is relatively high (Table 4 and Supl. Table 3). Both databases use different approaches to determining the language of a given work, and no published work to date has examined the quality of the language detection in OpenAlex.

Table 4: Spearman Correlation of Languages between OpenAlex and Scopus (2013-2022).

|  | **OpenAlex in Scopus** | **Scopus** |
|---|---|---|
| **Scopus** | 0.821 | - |
| **OpenAlex Total** | 0.954 | 0.772 |

Figure 8: Comparison of Number of Works by Language Between OpenAlex and Scopus.

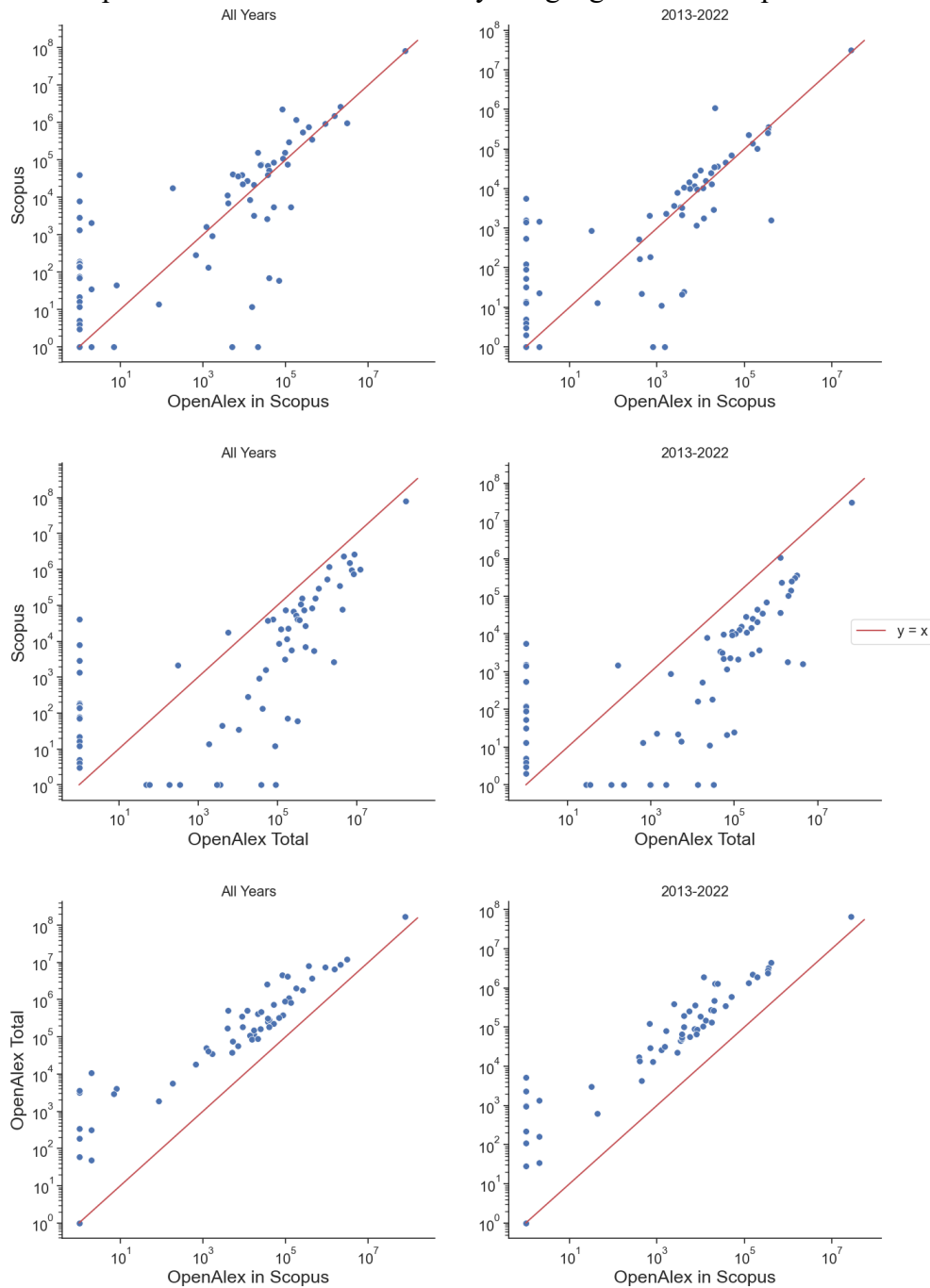

## 4. Discussion

This study shows that OpenAlex outperforms Scopus in coverage and provides evidence of the usefulness of OpenAlex for some forms of bibliometric analyses. However, it also indicates a need for further scrutiny of the accuracy and completeness of some important metadata elements, and points to areas that require further investigation before OpenAlex can be confidently used.

Our analysis shows that OpenAlex is a superset of Scopus and demonstrates what is gained by including sources not found in Scopus. Across the measures we examined, we find very high rank correlations between OpenAlex and Scopus. At the country level, when limited to the subset of works published in journals indexed by Scopus, we find that OpenAlex is also a reliable alternative. Including the remainder of OpenAlex has little effect on the rank correlations, although there are individual countries that move significantly in their position.

The lowest correlations were found for the number of works per language. OpenAlex does not have the same indexing restriction that Scopus does, limiting to works that have an English abstract and that should result in differences in works by language. However, we suspect discrepancies are also due to the differences in how each database identifies the language of the work: Scopus collects languages as a metadata attribute, whereas OpenAlex infers the language using a language detection algorithm on the title and abstract, which might lead to particular inaccuracies (especially when abstracts are available in multiple languages). Similarly, differences found between the number of works for each document type seem to largely stem from different classification systems used by each database. However, the use of the document type 'article' as a fall-back category, combined with the sizable number of articles of unknown source type in OpenAlex, suggests that this foundational metadata element deserves further scrutiny.

The ranking of countries by number of citations remains highly correlated, but unlike the analysis of the number of works, we find that, despite the significantly lower number of works, Scopus has more citations per country than OpenAlex. This follows from the high proportion of works with zero indexed references which results in a lower number of references per work, as compared to Scopus (Culbert et al., 2024).

Taken together, we draw two main conclusions: First, that for a limited set of analyses, OpenAlex can already be used as a replacement for traditional bibliographic databases. Our work offers insights into which data and which analyses can be relied upon and which require further scrutiny. Corollary to this, the second conclusion is that additional work is still needed to better understand the current limitations of OpenAlex and to improve its data quality and completeness. Doing so would lead to OpenAlex serving as a suitable and open replacement to traditional bibliographic databases and, thanks to its more inclusive indexing policies, to enable additional analyses not possible from a more constrained set of works.

To support these efforts, we conclude this paper with some suggestions for enhancements for OpenAlex and with a list of additional core questions that we believe the scientometrics community can answer to help the research community reliably use OpenAlex in the science of science and make further suggestions to OpenAlex on improvements needed.

## 5. Suggestions for further enhancements to OpenAlex

- As multiple studies have identified, affiliation data is often missing for a significant number of works. This data is crucial for many bibliometric analyses and so further efforts should be made to increase the completeness and accuracy of affiliation metadata, particularly for countries and institutions that have traditionally been less visible.
- OpenAlex still lags behind in the number of indexed references (and as a consequence, citations) when compared to Scopus. Indexing additional references will be necessary before citation analysis can take advantage of OpenAlex's extensive coverage. This would require a shift from relying on openly available citation data to actively extracting references from full texts and matching them to source items.

- More inclusive indexing offers the opportunity to better understand global contributions to scholarship, but there appears to be a need for better language detection algorithms (perhaps combined with the use of available language metadata).
- To guide and standardize analyses, OpenAlex could offer, or support the community to contribute, templates of filters that narrow in on commonly used subsets of documents (e.g., to remove retracted documents and paratext).
- We identified a large number of works for which the type of source could not be identified. Further disambiguation of this metadata is needed, along with syncing with source standards (e.g., ISSN).
- Content from Microsoft Academic Graph (MAG) makes up a sizable amount of the OpenAlex data. With MAG no longer receiving updates, an analysis of content that is only present in OpenAlex through the MAG corpus is needed, as are the following accommodations: i) identifying additional indexing strategies to capture content from those sources, or ii) eliminate material that is not really needed. Further, a filter could be added to easily include or exclude MAG-only content.
- Finally, one of the areas of greatest opportunity for OpenAlex lies in the indexing of works beyond journal articles. Expanding coverage and metadata quality of these document types would be beneficial.

## 6. Remaining Core Questions for the Community

- What is the recall and precision of key metadata fields (especially author and affiliation, document type, and language)? These calculations should be based on gold-standard datasets, such as those from author and institutional repositories or from other curated lists.
- What is the nature and extent of the coverage of non-article documents (e.g., books, research data, software, theses) currently found in OpenAlex? More broadly, what is still missing, what are known biases, and how can we as a community approach that question systematically?
- Classifying works by subject and delineating disciplines are central for field normalization and to identify interdisciplinarity. Traditional databases like Scopus and WoS rely on curated classification systems that determine a journal's or work's discipline. In contrast, OpenAlex concepts and topics are created algorithmically and have shown to exhibit numerous issues (Hug et al., 2017) that demand further scrutiny. How does this approach influence field-normalized citation indicators which are widely used in rankings exercises and research intelligence more broadly?
- How can OpenAlex best balance the inclusion of fields with diverse peer-review standards (STEM, health, social sciences, humanities, arts, law, business, etc) while limiting the inclusion of deceptive publishers (predatory journals and conferences)?
- While rank correlations of number of works between OpenAlex and other databases are high, what are the countries and institutions that gain the most? More broadly, what are the characteristics of the scholarship found in OpenAlex and not found in other databases?

**Supplementary Tables**

Supplementary Table 1. Comparison of Number of Works by Document Type Between OpenAlex and Scopus (All Years).

| Document Type | Scopus | OpenAlex | | | |
|---|---|---|---|---|---|
| | | Source In Scopus | Difference | All | Difference |
| article | 86,255,829 | 86,396,586 | 0% | 201,435,514 | 134% |
| book-chapter | 2,836,598 | 1,593,530 | -44% | 21,367,957 | 653% |
| other | 1,717 | 77,520 | 4415% | 12,566,103 | 731764% |
| book | 352,693 | 71,583 | -80% | 5,379,368 | 1425% |
| report | 13,915 | 11 | -100% | 822,915 | 5814% |
| editorial | 1,662,967 | 220,180 | -87% | 293,933 | -82% |
| erratum | 567,025 | 174,669 | -69% | 199,123 | -65% |
| letter | 2,269,458 | 43,567 | -98% | 72,835 | -97% |
| retractions | 26,728 | 0 | -100% | 0 | -100% |

Supplementary Table 2: Spearman Correlation of Countries between OpenAlex and Scopus (All Years).

| | OpenAlex in Scopus | Scopus |
|---|---|---|
| **Scopus** | 0.975 | - |
| **OpenAlex Total** | 0.995 | 0.969 |

Supplementary Table 3: Spearman Correlation of Languages between OpenAlex and Scopus (All Years).

| | OpenAlex in Scopus | Scopus |
|---|---|---|
| **Scopus** | 0.759 | - |
| **OpenAlex Total** | 0.956 | 0.753 |

**Open science practices**
OpenAlex data is openly available with a CC0 licence (OpenAlex, 2024). Scopus and SciVal data cannot be shared due to the proprietary nature of these databases, but details of the queries are provided in the methodology section for replication by others with access to these products.

Python code used to process and analyze data and to produce figures and tables is openly available under an MIT licence (Alperin, 2024).

## Acknowledgments

We would like to acknowledge the research support of Emily McGovern who collected citation data from SciVal. We would also like to thank Eric Schares for his helpful suggestions on earlier versions of this manuscript.

## Author contributions

J.P.: Conceptualization, Data curation, Formal analysis, Methodology, Software, Visualization, Writing - original draft, and Writing - review & editing. J.P.A.: Conceptualization, Data curation, Formal analysis, Methodology, Software, Visualization, Writing - original draft, and Writing - review & editing. K.D.: Conceptualization, Data curation, Formal analysis, Methodology, Writing - original draft, and Writing - review & editing. S.H.: Conceptualization, Methodology, Writing - original draft, and Writing - review & editing. V.L.: Writing - review & editing.

## Competing interests

Two of the authors of this study (JP & KD) are employees of OurResearch, the organisation that created and maintains OpenAlex.

## Funding information

This work was not supported by any external funding.